\documentstyle[12pt,a4wide]{article}
\input epsf
\newcommand{\be}{\begin{equation}}
\newcommand{\ee}{\end{equation}}

\newcommand{\I}{{\cal I}}
\newcommand{\bpi}{\mbox{\boldmath $\pi$}}
\newcommand{\pauli}{\mbox{\boldmath $\tau$}}
\font\mybb=msbm10 at 11pt

\def\bb#1{\hbox{\mybb#1}}

\def\bR {\bb{R}}

\renewcommand{\theequation}{\arabic{section}.\arabic{equation}}
\newcommand{\news}{\setcounter{equation}{0}}
\def\ben{\begin{equation}}
\def\een{\end{equation}}
\def\bea{\begin{eqnarray}}
\def\eea{\end{eqnarray}}
\begin{document}

\title{\vskip -0pt
\bf \large \bf 
SKYRMIONS AND THE 
{\mbox{\LARGE \boldmath $\alpha$}}-PARTICLE MODEL OF NUCLEI     
\\[30pt]
\author{Richard A. Battye$^{1}$, Nicholas S. Manton$^{2}$
and Paul M. Sutcliffe$^{3}$\\[10pt]
\\{\normalsize $^{1}$
{\sl Jodrell Bank Observatory, Macclesfield, Cheshire SK11 9DL, U.K.}}
\\{\normalsize {\sl $\&$  School of Physics and Astronomy,
Schuster Laboratory,}}
\\{\normalsize {\sl University of Manchester, Brunswick St,
 Manchester M13 9PL, U.K.}}
\\{\normalsize {\sl Email : rbattye@jb.man.ac.uk}}\\
\\{\normalsize $^{2}$  {\sl Department of Applied Mathematics and Theoretical
Physics,}}
\\{\normalsize {\sl
University of Cambridge, Wilberforce Road, Cambridge CB3 0WA, U.K.}}\\
{\normalsize{\sl Email : N.S.Manton@damtp.cam.ac.uk}}\\
\\{\normalsize $^{3}$  {\sl Institute of Mathematics,
University of Kent, Canterbury CT2 7NF, U.K.}}\\
{\normalsize{\sl Email : P.M.Sutcliffe@kent.ac.uk}}\\
{\normalsize {\sl Address from September 2006,}}\\
{\normalsize {\sl Department of Mathematical Sciences,
University of Durham, Durham DH1 3LE, U.K.}}\\
}}
\date{May 2006}
\maketitle
 
\begin{abstract}
We compute new solutions of the Skyrme model with massive pions.
Concentrating on baryon numbers which are a multiple of four, we
find low energy Skyrmion solutions which are composed of charge 
four sub-units, 
as in the $\alpha$-particle model of nuclei. We summarize our current 
understanding of these solutions, and discuss their relationship to
configurations in the $\alpha$-particle model.

\end{abstract}

\newpage

\section{Introduction}\news
Skyrme's vision was that the three pion fields form a nonlinear
semiclassical medium that dominates the interior of nuclei. He
proposed what is now called the Skyrme model \cite{Sk,Sk1}, whose
Lagrangian is a version of the nonlinear sigma
model, in which the pion fields $\bpi$ are combined into an 
$SU(2)$-valued scalar field
\be
U = (1 - \bpi \cdot \bpi)^{1/2}{ 1} + i \bpi \cdot \pauli \,,
\ee
where $\pauli$ are the Pauli matrices. There is an associated current,
with spatial components $R_i=(\partial_i U)U^\dagger$. Restricting to 
static fields, the energy in the Skyrme model is given by
\be
E=\frac{1}{12\pi^2}\int \left\{-{1 \over 2}\mbox{Tr}(R_iR_i)-{1 \over 16}
\mbox{Tr}([R_i,R_j][R_i,R_j])+m^2\mbox{Tr}(1-U)\right\} \, d^3x\,,
\label{skyenergy}
\ee
and the vacuum field is $U = 1$. In the above, the energy and length 
units (which must be fixed by comparison to real data) have been
scaled away, leaving only the pion mass parameter $m$. This parameter
is proportional to the (tree-level) pion mass, in scaled units.
Motivated by the results in Refs.\cite{BKS,BS11}, 
we shall set $m=1$ for most of our study. 

The model has a conserved, integer-valued topological charge
$B$, which is identified with baryon number. This is the degree of the
mapping $U: \bR^3 \to SU(2)$, which is well-defined, provided
$U \to 1$ at spatial infinity. In the above units
there is the Faddeev-Bogomolny energy bound, $E\ge |B|$.
Remarkably, for a purely pionic theory, Skyrme showed that 
there are topological soliton solutions -- Skyrmions -- that can be 
identified with baryons (for a review, see \cite{book}). The 
baryons are therefore self-sustaining structures in the 
pion field, and explicit nucleonic sources are not needed. 

Classically, the Skyrmions have no
spin, but they have rotational and isorotational collective 
coordinates which need to be quantized. In this way the basic, $B=1$
Skyrmion acquires spin and isospin, and its lowest energy states can 
be identified with spin $\frac{1}{2}$ protons and neutrons. The next
lowest states are identified with spin $\frac{3}{2}$ delta
resonances. The masses of the nucleons and deltas can be used to
calibrate the Skyrme model, and it is then found that other physical 
properties of quantized Skyrmions are in reasonable, but not excellent 
agreement with data \cite{ANW}.

Skyrme and later workers hoped that multi-Skyrmion solutions,
with baryon numbers greater than 1, could similarly be quantized and 
identified with nuclei. This programme has had 
some success \cite{BC,Ca,BTC,BS3,Walh,Ir}. Minimal energy solutions of the
Skyrme field equation, with baryon numbers $B = 2,3,4$ and $6$,
have the right properties to model the deuteron, the isospin pair 
$^3{\rm H}$ and $^3{\rm He}$, the $\alpha$-particle $^4{\rm He}$, and the 
nucleus $^6{\rm Li}$. In each case, the collective coordinate quantization is
constrained by the symmetries of the 
classical solution, and the need to impose Finkelstein-Rubinstein (FR) 
constraints \cite{FR}(which encode the requirement that the quantized $B = 1$
Skyrmion is a spin $\frac{1}{2}$ fermion). The resulting lowest
energy states for $B = 2,3,4$ and $6$ have spin/parity, respectively, 
$J^P=1^+, \frac{1}{2}^+, 0^+$ and $1^+$ (with isospin $T = 0$
in the even $B$ cases, and $T = \frac{1}{2}$ for $B=3$),
agreeing with those of real nuclei. However, in the standard
parametrization of the model the binding
energies are too large and the sizes too small. 

The classical solutions all have interesting shapes 
\cite{KS,Ma3,Ve,BTC,BS3};
they are not spherical like the basic $B = 1$ Skyrmion. The $B = 2$ 
Skyrmion is toroidal, and the $B = 3$ Skyrmion tetrahedral. The $B = 4$
Skyrmion is cubic and can be obtained by bringing together two $B = 2$ 
toroids along their common axis (with one flipped up-side down, which 
breaks the axial symmetry); finally, the $B = 6$ solution has 
$D_{4d}$ symmetry and can be formed from three toroids stacked one 
above the other. These structures, though very different from what one 
might expect from other models of nuclei, have some phenomenological 
support. Forest et al. \cite{Forest} have determined by Monte
Carlo methods the wavefunctions of a number of light nuclei, regarded as bound 
states of individual nucleons, and have shown that the tensor forces imply
that two-nucleon pairs predominantly form $T = 0$, $J = 1$ states 
(and exactly this for the deuteron) and that the spatial wavefunction
is concentrated in a toroidal region with the $z$-axis as symmetry axis  
for $J_z = 0$. The apparent dumbell shape that occurs for
$J_z={\pm}1$ can be interpreted as a toroid with its symmetry axis
in the $x$-$y$ plane, spinning about the $z$-axis. Furthermore,
pairs or triples of these tori, which occur in larger nuclei like 
${^4}\rm He$ and ${^6}\rm Li$, tend to have their axes lined
up, producing stacks of tori. 

So far there is no direct phenomenological
evidence for the hollow, cubically symmetric structure
of the $B = 4$ Skyrmion, but this may show up in a more refined
analysis of the nucleon position correlations in ${^4}\rm He$
wavefunctions, or possibly, by considering a superposition of $0^+$
and $4^+$ spin states \cite{MaMa}. 

Great progress has been made in determining the classical
solutions of the Skyrme model, at zero pion mass, for all baryon
numbers from $B=7$ up to $B = 22$ \cite{BS3}, and various larger values of 
$B$ \cite{BHS}. These solutions are all of the form of hollow
polyhedra, rather like carbon fullerenes. The baryon density is concentrated in
a shell of roughly constant thickness, surrounding a region whose
volume increases like $B^\frac{3}{2}$
 and where the energy and baryon density is
very small. Such a hollow structure clearly disagrees with the 
approximately constant baryon density observed in the interior of
real nuclei,
which implies that nuclear volume increases like $B.$
(The Skyrmions with $B$ between 2 and 6 also have zero baryon
density at the centre, but this is perhaps consistent with
the known dips of nuclear density seen in the cores of some small
nuclei, which are not easy to understand \cite{Anag}.) Another 
difficulty for the Skyrme model at larger $B$ is disagreement between the 
spin/parity assignments of the lowest energy states, and
those of real nuclei. In particular, for 
$B = 7$ the classical Skyrmion has a beautiful dodecahedral symmetry, 
but collective coordinate 
quantization leads to a lowest spin of $J=\frac{7}{2}$ (for 
$T = \frac{1}{2}$) \cite{Ir,Kr2}, disagreeing with the experimental values 
$J = \frac{3}{2}$ for the ground states of the isospin doublet 
${^7}\rm Li$ and ${^7}\rm Be$. This suggests that the
$B=7$ dodecahedral solution is too symmetric to be the ground state
and it would be preferable if a less symmetric solution existed,
which could have a larger classical energy than the dodecahedron,
but be quantized with a lower spin.

However, a crucial discovery has been made recently \cite{BS10,BS11}. 
For larger baryon numbers, $B\geq 8$, the value of the pion 
mass has an important qualitative effect.  The hollow
polyhedral solutions discussed above were obtained in the
limit where the pion mass vanishes. It is tempting to work in this
limit because the Skyrme model has greater symmetry 
there, and careful numerical work has shown that the classical 
solutions for $B$ less than about $8$ are fairly insensitive to a reduction 
of the pion mass from its physical value to zero \cite{BS10,HoMa2}. 
But it has now been established \cite{BS10,BS11} that the hollow 
polyhedral Skyrmions for $B \ge 8$ 
do not remain stable when the pion mass is set at a physically reasonable 
value \cite{AN,BKS}. The result, in retrospect, is not surprising. 
In the interior of the hollow polyhedra, the Skyrme field 
is very close to $U = -1$ (i.e. the antipode to the vacuum value), and
here the pion mass term gives the field a maximal potential energy. 
Consequently, the polyhedra become unstable to squashing modes. The 
interior region where $U$ is close to $-1$ tends to pinch off 
and separate into smaller sub-regions. 

New Skyrmion solutions have been found \cite{BS11} for $10\le B\le 16$
that have a planar character. They can be interpreted as bound arrangements 
of lower charge clusters in a planar layer, 
with $B=4$ cubes and $B=3$ tetrahedra often occuring.  
A cluster structure is an encouraging
development, since it is known since the 1930's that many nuclei 
with isospin zero and with $B$ a multiple 
of four may be described as arrangements of $\alpha$-particles \cite{BW}.
The planar Skyrmions found so far are local minima of the Skyrme
energy, but it is likely that there are a number of different local
minima, and it is not known which will be the global 
minima. It would be interesting if, over a larger range of baryon
numbers that are multiples of four, there exist Skyrmions which are 
composed of $B = 4$ sub-units, as in the $\alpha$-particle model of nuclei. 
In this paper we report the results of a search
for such solutions. We summarize our current 
understanding of these solutions, and discuss their relationship to
configurations in the $\alpha$-particle model.

Ideally, one would like to work with several values of the 
(scaled) pion mass, as the quantitative results are likely 
to be sensitive to its value. One should then
recalibrate the Skyrme model to fit, if possible, the masses and sizes 
of a few nuclei like $^{12}{\rm C}$ and $^{16}{\rm O}$, but 
this can only be done once one has the correct Skyrmion
solutions. (One would lose the fit to the delta resonance mass, but
this would be no great loss, as the delta resonance is broad, and
highly excited on the usual energy scale of nuclear states.)  
Unfortunately, it is computationally too expensive to perform numerical
simulations for a range of pion masses, so most of our results are
for $m=1$. However, the current evidence
\cite{BS11,HoMa2} suggests that 
the qualitative properties of a particular Skyrmion are not
very sensitive to variations in the pion mass over a range of values,
though as mass thresholds are crossed some stable solutions may cease
to exist and new local minima can appear.

We have used the same methods described in detail in Ref.\cite{BS3}
to numerically relax field configurations to static solutions of the
Skyrme model with $m=1.$  
Most of the results presented in this paper used grids containing
$100^3$ points with a lattice spacing $\Delta x=0.1,$ though larger grid sizes
and smaller lattice spacings were also used. The energy computations
in Ref.\cite{BS3} are for the Skyrme model with $m=0$ and appear to be 
accurate to within around $0.1\%.$ 
For $m=1$ there is a more rapid spatial 
variation of the fields and consequently we have found it more difficult
to accurately compute energies. We estimate that the errors could be
as large as $0.5\%,$ though we expect relative energies to be more 
accurate than this, as we discuss later.

\section{Skyrmions and {\mbox{\Large \boldmath $\alpha$}}-particles}\news
We shall now discuss Skyrmion solutions with baryon number $B$ a 
multiple of four, with the aim of making contact with the 
$\alpha$-particle model \cite{BW,BFWW}. The $\alpha$-particle 
model has considerable success 
describing the ground and excited states of nuclei with baryon number a 
multiple of 4, and isospin 0 (i.e. equal, even numbers of protons and 
neutrons). In the model, $\alpha$-particles are treated as point
particles subject to a phenomenological attractive potential
which becomes repulsive at short range, and in
their quantized states they form ``molecules''.

\subsection{$B=8.$}
We recall first that in the massless pion limit $m=0$, the $B=8$ Skyrmion is a
hollow polyhedron with $D_{6h}$ symmetry (a ring of 12 pentagons
capped by hexagons above and below), with no obvious relation to
a pair of cubic $B=4$ Skyrmions. However, motivated by the 
$\alpha$-particle model, we expect that at
$m=1$, the lowest energy solution is more likely to be a ``molecule'' of two
$B=4$ Skyrmions.

By considering the interaction of two $B=4$ Skyrmions
in the massless pion limit we will see why a connection 
with the $\alpha$-particle model is difficult to achieve in this case,
but the situation is improved by the introduction of massive pions.
With massless pions the leading order asymptotic fields of a $B=1$
Skyrmion resemble a triplet of orthogonal pion dipoles, and hence
the interactions of well-separated Skyrmions can be understood in
terms of scalar dipole-dipole forces \cite{book}. 
The $B=4$ cube has no dipoles, only a doublet of quadrupoles \cite{Ma4}, 
so two $B=4$ cubes interact rather weakly by Skyrme 
model standards. Two $B = 4$ cubes placed in the same 
orientation and next to each other weakly attract, as can be
confirmed by calculating the quadrupole-quadrupole interaction.
There is an associated Skyrme configuration, but it is only a 
saddle point, not a local energy minimum. Because of a significant short-range 
octupole interaction in the single pion field component that has no
quadrupole moment, it is better to twist one cube by $90^0$ 
relative to the other around the axis joining them. Such an initial
configuration can be constructed using Skyrme's product ansatz (see Appendix), 
and is presented in Fig.~\ref{fig-8}A. With 
the stronger attraction the cubes now merge, resulting in the stable 
solution displayed in Fig.~\ref{fig-8}B. This is a hollow polyhedral
Skyrmion which has cubic symmetry $O_h$, and the associated polyhedron
is the truncated octahedron, a figure with 14 faces (8 hexagons on the
vertices of a cube and 6 squares on the faces)
whose face centres correspond to the holes in the Skyrmion.

\begin{figure}[ht]
\begin{center}
\leavevmode
\vskip -0cm
\epsfxsize=12cm\epsffile{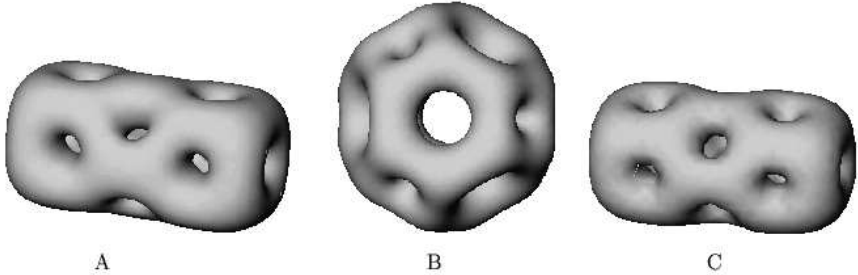}
\caption{Baryon density isosurfaces; A) product ansatz of two
$B=4$ cubes with a spatial rotation of one of the cubes through $90^\circ$
around the line joining the cubes; B) the $B=8$ truncated octahedron 
with $m=0$; 
C) the relaxed $B=8$ Skyrmion with $m=1$ obtained from perturbing
the truncated octahedron.}
\label{fig-8}
\vskip 0cm
\end{center}
\end{figure}

The truncated octahedron solution can be constructed 
using the rational map ansatz 
that was introduced in Ref.\cite{HMS} (and is reviewed in the Appendix), 
which has been found to be a most effective 
analytical approximation for all the hollow polyhedral
solutions at zero pion mass. The rational map for a $B = 8$
truncated octahedron field is
\be
R(z) = \frac{z^2(z^4 + 1)}{z^8 - 10z^4 + 1} \,.
\ee
Together with a suitable radial profile, this gives a starting ansatz
for the Skyrme field $U$. The true solution is obtained by relaxing
this, while preserving $O_h$ symmetry. For $m=0$ it is a 
low energy local minimum, slightly higher in energy than the Skyrmion
with $D_{6d}$ symmetry. 

If we now consider this truncated octahedron Skyrmion 
as an initial configuration
in the Skyrme model with pion mass $m=1$ then we find that it is unstable.
A small squashing, along an axis not aligned with any of the cubic symmetry
axes, followed by a numerical relaxation, 
produces the stable solution displayed
in Fig.~\ref{fig-8}C, which has $D_{4h}$ symmetry. In the process, 
the central $8$-fold degenerate point where $U = -1$ splits
symmetrically into two clusters near the centres of the two cubes. 
(It appears that in the middle of each cube,
$U = -1$ occurs with $2$-fold degeneracy at a pair of slightly
separated points.)

The new solution is 
clearly a bound configuration of two $B = 4$ cubic solutions, and closely
resembles the configuration shown in Fig.~\ref{fig-8}A.
This matches the known physics that ${^8}\rm Be$ is an almost bound state 
of two $\alpha$-particles. The spectrum of ${^8}\rm Be$ resonances is 
well enough known that this nucleus can be treated as a molecule of two 
$\alpha$-particles in an attractive potential, not quite strong enough to 
produce a bound state. The $B = 8$ Skyrmion should be thought of as 
the classical solution corresponding to the minimum of the potential. 
The classical energy required to break the solution into two
well-separated $B = 4$ clusters is small. If kinetic and Coulomb effects
were included, the solution might naturally unbind.
The FR constraints allow the lowest quantum state for this 
solution to be a $0^+$ state, with zero isospin, consistent with the 
quantum numbers of ${^8}\rm Be$. 

Note that in Ref.\cite{BS11} it was shown that taking as an initial
condition the $m=0$ minimal energy $D_{6d}$ polyhedral Skyrmion and
relaxing it (after a small symmetry breaking perturbation) 
in the Skyrme model with $m=1$ yields essentially the same polyhedral
solution back again. The $D_{6d}$ symmetry is restored and the only
significant change is a small squashing in the direction of the main
symmetry axis. Thus we have found two very different local minima
in the massive pion model, and there could be others. 

For $m=1$ 
we compute that, to three decimal places, the energies per baryon of the 
$D_{6d}$ and new $D_{4h}$ solution are both equal to  
$E/B=1.294.$ For comparison, the energy per baryon of the $B=4$ cube is
$E/B=1.307$ for $m=1$. 
These values were obtained on a grid containing $100^3$
lattice points with a lattice spacing $\Delta x=0.1.$ We have also computed
the energies of these solutions on a grid with $200^3$ points with
$\Delta x=0.05$ and obtain energies of $E/B=1.297$ for both solutions.
Thus we estimate that numerical errors could be as large as $0.5\%$ 
in our energy computations,
though we expect relative energies to be more accurate than this.
To four decimal places, on the grid with $200^3$ lattice points, 
we found $E/B=1.2970$ and $E/B=1.2967$, for the $D_{6d}$ and
$D_{4h}$ solutions respectively. This suggests that the $D_{4h}$
solution might have the lower energy but as the energy difference
is much less than the numerical
accuracy to which we can reliably compute, we are unable to make
any conclusive statement about which (if either) of these solutions is
the global minimum.

\subsection{$B=12$}

In the $\alpha$-particle model 
the classical minima of the potential energy for three or
four $\alpha$-particles occur, respectively, for an equilateral
triangle and a regular tetrahedron \cite{BW,BFWW,WBFF}. In this and the 
following subsection we shall investigate whether Skyrmion analogues 
of these, and other, results can be found.  

\begin{figure}[ht]
\begin{center}
\leavevmode
\vskip -0cm
\epsfxsize=8cm\epsffile{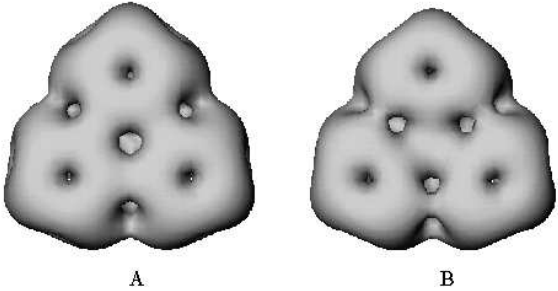}
\caption{Top and bottom views of the baryon density isosurface of 
a $B=12$ Skyrmion with triangular symmetry, formed from the product
ansatz of three cubes in a triangle.}
\label{fig-12tri}
\vskip 0cm
\end{center}
\end{figure}

Motivated by the $\alpha$-particle model of $^{12}{\rm C}$,
we have sought a triangular $B=12$ solution in the Skyrme model, composed 
of three $B=4$ cubes, oriented so that they attract. 
A configuration with approximate $D_{3h}$ symmetry can be obtained
by using the product ansatz to place cubes with their centres on the 
vertices of an equilateral triangle, and meeting at a common edge. 
Each cube is related to its neighbour by a spatial rotation
through $120^\circ$ followed by an isorotation by $120^\circ$. The
isorotation cyclically permutes the values of the pion fields on the 
faces of the cube, so that these values match on touching faces. 
This produces a field in which each pair of cubes is a 
slightly bent version of the new $B = 8$ solution described above.
However, it is fairly easy to see that around the centre of the triangle 
the field has a winding equivalent to a single Skyrmion, and it
is unstable to a perturbation (included automatically by the product ansatz)
that breaks the up-down reflection symmetry in the plane of the triangle.
The instability results, upon relaxation, 
in the single Skyrmion at the centre of the triangle
moving down and merging with the bottom face of the triangle; in fact it
fills what was a hole in the baryon density isosurface. Apart from this 
difference, the resulting relaxed solution, which has only a $C_3$ symmetry,
is very similar to the initial field formed from the three cubes.
This $C_3$ symmetric Skyrmion is presented in Fig.~\ref{fig-12tri}, 
where views from both the top and bottom are displayed, so that the up-down
symmetry breaking is clearly demonstrated. We compute that $E/B=1.288.$

The top view in Fig.~\ref{fig-12tri} is very similar to the minimal
energy $B=11$ Skyrmion with $m=0$ \cite{BS3}, and suggests that the
initial arrangement of three cubes with an approximate $D_{3h}$
symmetry can also be viewed as the $m=1$ version of this $B=11$ Skyrmion
with a single Skyrmion placed inside at the origin. Such a field configuration
can be constructed with exact $D_{3h}$ symmetry using the 
  double rational map ansatz \cite{MP} (see
Appendix). This involves a $D_{3h}$ symmetric
 outer map of degree 
11, $R^{\rm out}$, and a spherically symmetric 
degree 1 inner map, $R^{\rm in}$, together with an overall radial 
profile function. Explicitly the maps are
\begin{eqnarray}
	R^{\rm out} &=& \frac{z^2(1+az^3+bz^6+cz^9)}{c+bz^3+az^6+z^9}
 \label{map09}\\
	R^{\rm in} &=& -\frac{1}{z} \,, \label{map01}
\end{eqnarray}
where the real constants
are $a=-2.47$, $b=-0.84$ and $c=-0.13.$ Note that the  orientation
of the inner map has to be chosen so that the $D_{3h}$ subgroup
is realized in a way compatible with the degree 11 map.
Numerically relaxing the field of the double rational map ansatz for $U$
yields a solution which is three cubes with exact  $D_{3h}$ symmetry,
but as we discussed above, this is only a saddle point and not a local
energy minimum.

In Ref.\cite{BS11} a $B=12$ solution with $C_3$ symmetry
was found, but this is different from the $C_3$ symmetric solution
described above.  The solution of 
Ref.\cite{BS11} is composed of three $B=3$ tetrahedra on the
vertices of an equilateral triangle and three single Skyrmions on the
vertices of the dual triangle. Its energy was found to be $E/B=1.289.$
Once again the energies of these two different local minima are too close
to have any confidence which of the two solutions has the true lowest energy.
It is a general observation of this work that rearrangements of clusters
have only a tiny effect on the energy of a Skyrmion, so as $B$ increases 
one expects an increasingly large number of local minima with extremely close
energies.

\begin{figure}[ht]
\begin{center}
\leavevmode
\vskip -0cm
\epsfxsize=6cm\epsffile{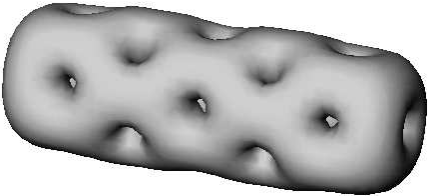}
\caption{$B=12$ Skyrmion formed from three cubes in a line, with the middle
cube being rotated by $90^\circ$ around the line of the cubes.}
\label{fig-12line}
\vskip 0cm
\end{center}
\end{figure}

Rearranged solutions are analogous to the
rearrangements of the $\alpha$-particles which model excited states of 
nuclei. An example is the Skyrme model analogue 
of three $\alpha$-particles in a chain configuration for the 7.65 MeV excited 
state of $^{12}{\rm C}$, with spin/parity $0^+$ \cite{Mor,FSW}.
We have found a suitable stable local minimum to describe three 
$\alpha$-particles in a chain. It is a generalization of the $B=8$ Skyrmion
and consists of three cubes placed next to each other in a line. 
The two end cubes have the same orientation, but the middle cube 
is rotated relative to the other two by $90^\circ$ around the axis of 
the chain. 
This $B=12$ Skyrmion is displayed in Fig.~\ref{fig-12line}, and has 
energy $E/B=1.285$. Thus it appears to have the lowest energy of 
the $B=12$ solutions we have found, but this is not certain given 
our numerical accuracy. Note that the energy difference we are trying
to understand, 7.65 MeV, is less than 0.1\% of the total energy of a 
$^{12}{\rm C}$ nucleus, and is therefore smaller than the 
present accuracy of our numerical energy computations.

To conclude, for $B=12$ we have found two new solutions in addition to the one
already known from Ref.\cite{BS11}. These new solutions are similar to those
found in the alpha particle model and all three have equal energy within
the level of accuracy we can achieve.

\subsection{$B=16$}
\begin{figure}[ht]
\begin{center}
\leavevmode
\vskip -0cm
\epsfxsize=16cm\epsffile{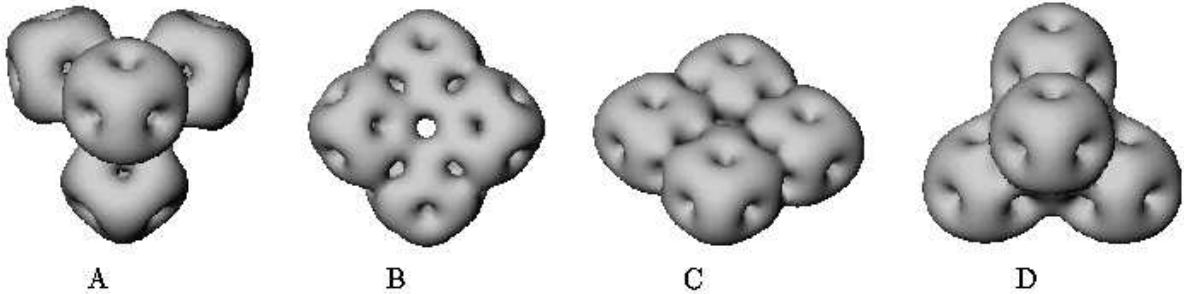}
\caption{$B=16$ Skyrmions composed of four cubes; A) tetrahedral arrangement;
 B) bent square; C) flat square; D) another tetrahedral arrangement.}
\label{fig-16}
\vskip 0cm
\end{center}
\end{figure}

We have found a tetrahedrally symmetric $B=16$ solution which is 
an arrangement of four $B = 4$ cubes. It was created using rational
maps to provide a tetrahedrally symmetric initial condition.
One again needs to use the double rational map ansatz 
as a starting point. This involves an outer map of degree 
12, $R^{\rm out}$, and an inner map of degree 4, $R^{\rm in}$, with compatible 
symmetries, and an overall radial profile function. Essentially one 
is filling a hollow $B = 12$ Skyrmion with 
a $B = 4$ Skyrmion. There is a $T_d$ symmetric map
which approximates the $m=0$ Skyrmion with $B = 12$ \cite{BS3},
and this can be combined with the $O_h$ symmetric map familiar 
from the $B = 4$ Skyrmion, giving $T_d$ symmetry overall. The maps are
\begin{eqnarray}
	R^{\rm out} &=& \frac{ap_+^3 + bp_-^3}{p_+^2 p_-} \label{map12}\\
	R^{\rm in} &=& \frac{p_+}{p_-} \,, \label{map4}
\end{eqnarray}
where $p_\pm(z) = z^4 \pm  2\sqrt{3}\,iz^2 + 1$, and the real constants
are $a=-0.53$ and $b=0.78$.
Starting from the double rational map ansatz for $U$, one lets the
field relax numerically, preserving the $T_d$ symmetry. The result is
the solution displayed in Fig.~\ref{fig-16}A. Initially $U = -1$ 
on a whole spherical surface, but after relaxation, $U = -1$ occurs at
16 points, clustered into groups of four points close to the centre 
of each cube.

The energy of this solution is $E/B=1.288,$ and the four cubes
are clearly visible in a tetrahedral arrangement. The cubes  
are surprisingly distinct, in comparison with the earlier
solutions in which cubes merge to a reasonable extent.  

This tetrahedral solution is only a saddle point. A perturbation which
breaks the tetrahedral symmetry, followed by a numerical relaxation,
reveals that it is energetically more favourable for the two cubes on 
a pair of opposite edges of the tetrahedron to open out, leading to 
the $D_{2d}$ symmetric solution presented in Fig.~\ref{fig-16}B. 
This Skyrmion resembles 
a bent square with four cubes on the vertices, and has a slightly lower 
energy, 
 $E/B=1.284$. Possibly, Coulomb effects could return the tetrahedral solution 
to stability, and there is always the possibility that the energy
ordering of these two competing configurations could be reversed
for an increased value of the pion mass parameter $m$; however, an
initial investigation reveals that this last possibility does not happen 
for $m=2$.  

A stable tetrahedral solution would be preferable. It has been shown that 
the closed shell structure of $^{16}{\rm O}$, just slightly perturbed, 
is compatible with clustering into a tetrahedral arrangement of four 
$\alpha$-particles. Moreover, the ground state and 
the excited states at 6.1 MeV and 10.4 MeV, with spin/parity $0^+$, $3^-$ 
and $4^+$, and some higher states, look convincingly like a rotational 
band for a tetrahedral intrinsic structure \cite{Den,Rob}.

It is interesting to note that a configuration similar to our
bent square Fig.~\ref{fig-16}B has been found as a low energy
intrinsic state using an $\alpha$-cluster model, where it is
termed a bent rhomb \cite{BauSS}. Furthermore, when Coulomb effects
are not included it has been found that the regular tetrahedron
is not the ground state, but rather an elongated tetrahedron is
preferred \cite{AbIr}. Thus there is some support, using more traditional
nuclear models, for the qualitative features of the solutions we have found.  

There is a further solution of low
energy, in which four $B=4$ cubes all have the same orientation, and
are connected together to form a flat square (see Fig.~\ref{fig-16}C).
This solution has been obtained by using an initial condition derived from
the product ansatz of four cubes, and also from a very different initial
condition constructed using a rational map with $D_{4h}$ symmetry.
The solution has energy $E/B=1.293$, so one might expect that it is
only a saddle point, having an unstable mode that bends the square to
the solution of Fig.~\ref{fig-16}B. However, perturbations of this
solution have failed to excite such a mode, so the current evidence suggests
that it may be a local minimum. 

Yet another tetrahedral configuration of four cubes exists. It is obtained
from an initial condition using the single rational map
$\tilde R(z) = c R(z)^4$, where $c$ is an arbitrary positive real 
parameter and $R$ is the cubic map (\ref{map4}) of degree four. If $c=1$ 
then this map has cubic symmetry $O_h$, but for all other values of $c$ the
symmetry reduces to tetrahedral, $T_d$. Relaxing such a tetrahedral
initial condition preserves the symmetry and yields the solution
shown in Fig.~\ref{fig-16}D. It is again a tetrahedral arrangement of
four cubes, but with different spatial orientations of the cubes than
in the earlier solution Fig.~\ref{fig-16}A. This solution resembles
half of the $B=32$ crystal chunk (discussed later in Section \ref{sec-32})
where alternate cubes have been removed. Its energy
is $E/B=1.295$ and therefore higher than the earlier tetrahedral
solution. A perturbation that breaks the tetrahedral symmetry again
results in the bent square solution of Fig.~\ref{fig-16}B.

In Ref.\cite{BS11} a planar $B=16$ solution was found that is not
composed of cubes and has an energy $E/B=1.288.$ This energy is very slightly
higher than that of the bent square $E/B=1.284,$ so our current belief
is that for $m=1$ the bent square may be the global minimal 
energy Skyrmion with $B=16.$

In summary, we have found four new $B=16$ solutions of the Skyrme
model with $m=1.$ Two of these have tetrahedral symmetry and are saddle
points, and the other two appear to be local minima, with the bent
square being a good candidate for the global minimum energy Skyrmion. 

\subsection{$B=20$}
\begin{figure}[ht]
\begin{center}
\leavevmode
\vskip -0cm
\epsfxsize=9cm\epsffile{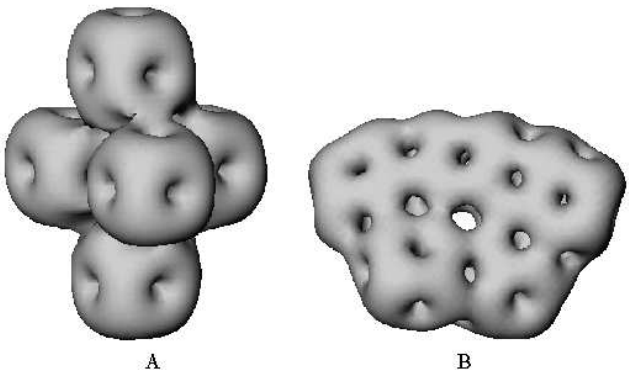}
\caption{A) Initial configuration of five cubes in a
triangular bipyramid arrangement; B) the final relaxed $B=20$ Skyrmion,
which is a more planar arrangement of five deformed cubes.}
\label{fig-20}
\vskip 0cm
\end{center}
\end{figure}
In the $\alpha$-particle model, 
the classical minimum of the potential energy for five $\alpha$-particles 
is a triangular bipyramid. An initial condition of five
$B=4$ cubes, with the same spatial and isospin orientations, placed on the
vertices of a triangular bipyramid is displayed in Fig.~\ref{fig-20}A.
The relaxation of this starting configuration completely changes the
shape and produces the solution presented in Fig.~\ref{fig-20}B.
This solution is still composed of five cubes, but they are substantially
deformed and create a more planar arrangement. It has a
$C_2$ symmetry and each cube is twisted slightly compared to its
neighbours, in a manner similar to that of the $B=16$ bent square.
The energy is $E/B=1.283$.

\begin{figure}[ht]
\begin{center}
\leavevmode
\vskip -0cm
\epsfxsize=9cm\epsffile{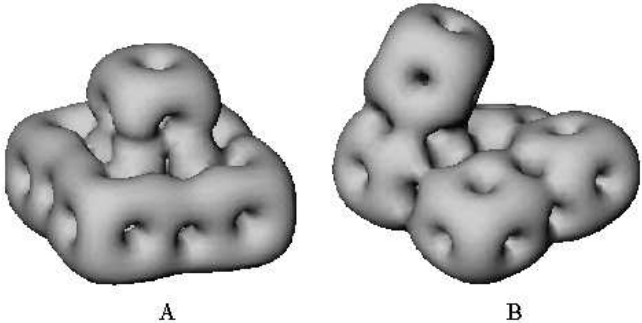}
\caption{A) Initial configuration of four cubes in a square
with a fifth cube placed on top of the square and in the centre; 
B) the final relaxed $B=20$ Skyrmion, with four cubes in a bent
square and a fifth cube joined to one of the corners of the bent square.}
\label{fig-20B}
\vskip 0cm
\end{center}
\end{figure}

Motivated by the results for $B=24$, which are presented in the following 
section, we consider an initial condition (see Fig.~\ref{fig-20B}A) 
consisting of four cubes in a square
with a fifth cube placed on top of the square and in the centre. All five
cubes have the same spatial orientation and the four cubes in the square
have the same isospin orientation, but the fifth cube has been given 
an isospin rotation by $180^\circ.$ This isospin rotation aids the attraction
 between the cubes and is motivated by trying to match the pion fields
of the fifth cube with the pion fields at the centre of the square of
cubes. Relaxation of this initial configuration produces the solution
shown in Fig.~\ref{fig-20B}B. It consists of the
$B=16$ bent square Skyrmion of Fig.~\ref{fig-16}B, together with an
extra cube joined to one of the corners of the bent square. 
This solution has energy $E/B=1.285,$ again demonstrating that the 
rearrangment of cubes has only a tiny effect on the energy. 

\subsection{$B=24$}
\begin{figure}[ht]
\begin{center}
\leavevmode
\vskip -0cm
\epsfxsize=9cm\epsffile{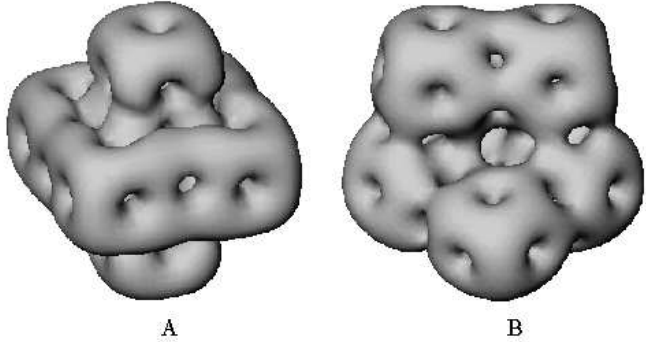}
\caption{A) Initial configuration of six cubes in a square bipyramid 
arrangement; B) the final relaxed $B=24$ Skyrmion,
which is four cubes in a bent square with two further cubes along the
diagonal of the top of the bent square.}
\label{fig-24}
\vskip 0cm
\end{center}
\end{figure}
The results we have found so far with $B=16$ and $B=20$ suggest that
the classical minimum of the potential energy for point $\alpha$-particles 
is not a good guide to predicting the cluster arrangements of $B=4$ Skyrmions
in energy minimizing solutions. Nonetheless, the point particle approximation
does suggest reasonable starting configurations, which can be created using
the product ansatz, ensuring that any unwanted symmetry is only approximate
and can therefore be destroyed by the relaxation process. 
For $B=24$ we begin with six $B=4$ cubes on the vertices of an octahedron,
but all with the same spatial orientation, so there is no octahedral symmetry
even approximately (see Fig.~\ref{fig-24}A). 
The two cubes above and below the four in a square
have been given an isospin rotation by $180^\circ,$ as this aids the 
attraction between the cubes. As in the previous section, the relaxed
solution, shown in  Fig.~\ref{fig-24}B, 
is still formed from $B=4$ sub-units, but it is very different
from the initial condition. It can clearly be seen that this $B=24$ solution
is the $B=16$ bent square Skyrmion of Fig.~\ref{fig-16}B, joined with the $B=8$
Skyrmion of Fig.~\ref{fig-8}C. The two cubes of the $B=8$ Skyrmion lie
across the diagonal of the $B=16$ bent square, so a $C_2$ rotational
symmetry is preserved. The fact that these $B=16$ and $B=8$ solutions
appear as sub-structures in the $B=24$ solution is further evidence that
these Skyrmions are the minimal energy arrangements of $B=4$ cubes.
The energy of this $B=24$ Skyrmion is $E/B=1.282.$

\subsection{$B=28$}\label{sec-28}
\begin{figure}[ht]
\begin{center}
\leavevmode
\vskip -0cm
\epsfxsize=9cm\epsffile{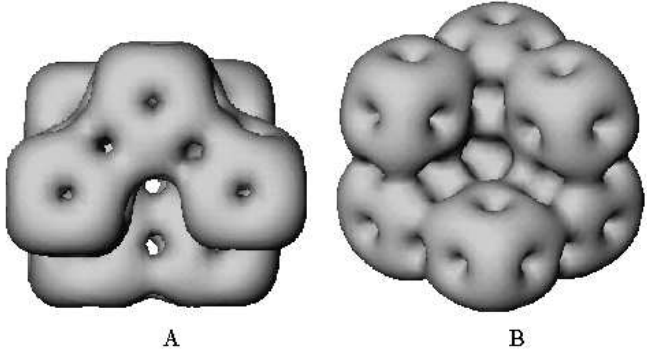}
\caption{A) Initial configuration of four cubes in a square with
three extra cubes placed above the square but not aligned with
the cubes below; B) the final relaxed $B=28$ Skyrmion,
which resembles the cubic $B=32$ Skyrmion but with one cube removed.}
\label{fig-28}
\vskip 0cm
\end{center}
\end{figure}

For both $B=20$ and $B=24$ we have found low energy solutions which contain the
$B=16$ bent square as a sub-structure. It therefore seems likely that a 
similar solution might exist for $B=28.$ In an attempt to find such a solution
we used the starting configuration shown in Fig.~\ref{fig-28}A, with four
cubes in a square and three extra cubes placed above the square, but directly
over holes in the square that do not correspond to faces of the cubes
below. All the cubes have the same space and isospace orientations. 
We do not expect this to be a very attractive arrangement, and indeed
this is one of the motivations for choosing this initial condition, since we
expect the cubes to rearrange into a low energy solution. 

Fig.~\ref{fig-28}B
displays the solution which results from the relaxation. Note that
each of the three cubes on top of the square has indeed aligned with a cube
in the four below, but that these four cubes have not formed the bent
square of Fig.~\ref{fig-16}B but rather remain in the flat square of
 Fig.~\ref{fig-16}C. The presence of the three connected cubes above appears
to suppress the tendency of the square of cubes to bend. 
The cubic $B=32$ Skyrmion (discussed in detail in the next section)
has eight cubes on the vertices of a larger cube and this 
$B=28$ solution clearly resembles the $B=32$ Skyrmion with one 
of the eight cubes removed. The energy of this solution is 
$E/B=1.279,$ which is slightly lower than might have been expected
given the previous energies for $B=16,20,24.$ This solution therefore
appears to be a a good candidate for the minimal energy $B=28$ Skyrmion
with $m=1.$ 

\subsection{$B=32$}\label{sec-32}
In Ref.\cite{BS11} it was shown that, even for relatively small values of the 
pion mass parameter $m$, the energy of a $B=32$ cubic Skyrmion is
lower than that of the minimal energy, 
hollow polyhedral fullerene-type Skyrmion.
This makes the $B=32$ cubic Skyrmion a candidate for the minimal energy
Skyrmion at this baryon number. It may be thought of as eight $B=4$ cubic 
Skyrmions placed on the vertices of a cube, each with the same spatial 
and isospin orientations. Alternatively, it may be created by cutting out
a cubic $B=32$ chunk from the infinite, triply-periodic Skyrme crystal
\cite{Ba}. However, both these constructions appear a little artificial
in that the final solution does not differ greatly from the initial
condition. 

\begin{figure}[ht]
\begin{center}
\leavevmode
\vskip -0cm
\epsfxsize=9cm\epsffile{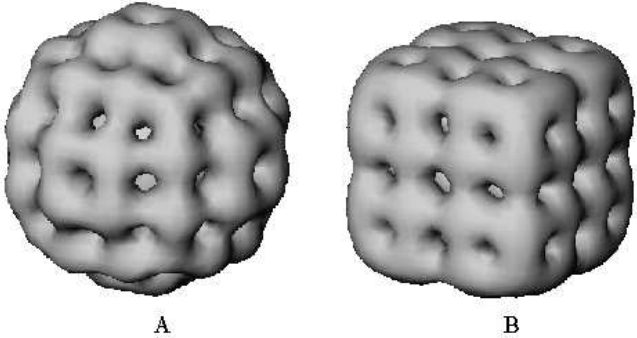}
\caption{A) Initial condition of the cubic $B=4$ Skyrmion inside
a cubic $B=28$ Skyrmion; B) the final relaxed $B=32$ configuration,
which is a chunk of the Skyrme crystal.} 
\label{fig-32}
\vskip 0cm
\end{center}
\end{figure}

A more convincing method to obtain the $B=32$ crystal chunk is to begin
with an initial condition given by the double rational map ansatz. We place
a $B=4$ cube inside a $B=28$ Skyrmion with cubic symmetry using the maps
\begin{eqnarray}
	R^{\rm out} &=& \frac{p_+(ap_+^6 + bp_+^3p_-^3-p_-^6)}
{p_-(p_+^6-bp_+^3p_-^3-ap_-^6)} \label{map28}\\
	R^{\rm in} &=& \frac{p_+}{p_-} \,, \label{map4again}
\end{eqnarray} 
where $a=0.33$ and $b=1.64$, and $p_\pm(z)$ are as before. The 
initial condition is displayed in
Fig.~\ref{fig-32}A. The numerical relaxation yields the solution presented 
in Fig.~\ref{fig-32}B, which is indeed the $B=32$ crystal chunk, with 
energy $E/B=1.274$. We have also
performed a similar simulation in which the initial double rational map
Skyrme field is perturbed to break the cubic symmetry, and the result
is again the cubic $B=32$ solution, suggesting it is a stable
local minimum. Furthermore, the cubic $B=32$ solution also arises 
from a simulation in which the initial conditions were created using the 
product ansatz to place eight cubes in an arrangement where seven of
the cubes are as shown in Fig.~\ref{fig-28}A, and the eighth cube is placed
in the obvious missing location to create a cross of four cubes. 

Note that slicing the $B=32$ crystal chunk in half produces the square
$B=16$ solution of Fig.~\ref{fig-16}C, whereas if one removes every alternate
cube then the result is the tetrahedral $B=16$ solution of 
Fig.~\ref{fig-16}D. As we remarked in the previous section, removing
a single cube produces the $B=28$ solution of  Fig.~\ref{fig-28}B.

Recall that a Skyrme field created using the original rational map ansatz 
with a degree $B$ map has a polyhedral structure with $2B-2$ holes in 
the baryon density \cite{HMS}. As can easily be seen from 
Fig.~\ref{fig-32}B the $B=32$ crystal chunk contains 54 exterior holes
and this corresponds to a degree 28 map. This was one of the main motivations
for considering the above double rational map construction using maps
of degrees 28 and 4. The next crystal chunk with cubic symmetry contains
27 cubes and therefore has $B=108.$ To create this from a triple 
rational map ansatz, by wrapping a third shell around the $B=4$ and $B=28$
shells, would therefore require a map with degree $B=76.$ A degree 76
map has 150 holes and it is easy to see that this is precisely the correct
number of exterior holes for the $B=108$ crystal chunk. It therefore seems
likely that a triple rational map ansatz exists that would provide suitable
initial conditions, with exact cubic symmetry, which relaxes to the $B=108$
crystal chunk. However, there is a 6-parameter family of $O_h$ symmetric
degree 76 maps and it is not clear how to obtain an appropriate member of
this family with a suitable distribution of the 150 holes. In the simpler 
case of the required degree 28 cubic map it turns out that the energy 
minimizing map is suitable, but we have verified that for the degree 76
map the energy minimizing map does not have the required distribution
of holes.    

\subsection{$B=7$}
$B=7$ Skyrme fields with cluster sub-units also look interesting. 
A $D_{3d}$ symmetric deformation of small energy of the $B = 7$ 
dodecahedral Skyrmion solution \cite{BS3}
looks like a pair of $B = 4$ cubes, sharing one $B = 1$ 
Skyrmion, and it should be possible to quantize this with a lower spin 
than $J=\frac{7}{2}$. A further, less symmetric deformation should 
split this into two clusters, a $B = 4$ cube and a $B = 3$ tetrahedron. It is 
also likely that if a $B = 1$ Skyrmion collides (classically) with this 
field configuration the resulting structure will break into two $B = 4$
cubic Skyrmions, thereby modelling the reaction 
${\rm p}$ + $^{7}{\rm Li}$ $\rightarrow$ $^{4}{\rm He}$ + $^{4}{\rm He}$.

We have attempted to construct a stable $B=7$ Skyrmion solution which
is less symmetric than the dodecahedral solution, for example by
using initial conditions containing a $B=3$ tetrahedron and a $B=4$
cube, but so far all our attempts have failed, in that the dodecahedral
solution is always recovered.

\section{Conclusion}\news
This work has shown that the Skyrme model with 
positive pion mass has qualitatively new solutions for baryon numbers 
$B\geq 8$. These are not hollow polyhedra, but more dense structures 
with clear clustering into $B = 4$ cubes, the Skyrme model
analogue of $\alpha$-particles. This gives confidence that the model 
is a true competitor to other successful models of nuclei. New solutions 
for $B = 8, 12, 16, 20, 24$ and $28$, and also the crystal chunk for 
$B = 32$ discovered 
earlier, which is a cubic arrangement of eight $B = 4$ cubes, are the 
Skyrme model analogues of $\alpha$-particle ``molecules''. Their 
quantization should give sensible rotational bands. 
Methods that may be useful in the quantization of these solutions
have been developed recently \cite{Kr3}. 
We have found that 
the cubic $B = 4$ Skyrmions can be rearranged with quite small energy 
changes. These rearranged solutions are analogous to the
rearrangements of the $\alpha$-particles which model excited states of 
nuclei.

Greater numerical precision would be helpful to determine the Skyrmion 
energies, and the energy gaps between nearby solutions. We hope then 
to recalibrate the Skyrme model's three parameters by fitting to, 
say, the $^{12}{\rm C}$ and $^{16}{\rm O}$ masses and sizes, together
perhaps with the excitation energy of the lowest $3^-$ state in 
$^{16}{\rm O}$. We expect the parameters to be 
considerably different from those which emerge by fitting the 
proton mass, proton size, and proton-delta mass difference, but believe 
they will be more suitable for modelling nuclei.

\renewcommand{\theequation}{A.\arabic{equation}}
\section*{Appendix: Product and Rational Map Ans\"atze}\news

Here, we briefly review the product ansatz \cite{Sk1}, 
the rational map ansatz \cite{HMS},
and the double rational map ansatz \cite{MP} for Skyrmions. These give
approximate, qualitatively useful field configurations, but not exact
solutions of the Skyrme field equation. 

The product ansatz is quite simple. Given $SU(2)$-valued Skyrme 
fields (often Skyrmion solutions) $U_1(\bf x)$ with baryon number 
$B_1$ and $U_2(\bf x)$ with baryon number $B_2$, one multiplies them, 
obtaining $U({\bf x}) = U_1({\bf x})U_2({\bf x})$. The field $U$ 
has baryon number $B_1 + B_2$.

If the baryon densities of $U_1$ and $U_2$ do not overlap significantly,
then $U$ is a useful superposition of the two field
configurations. The difference between the energy of $U$, and the
sum of the energies of $U_1$ and $U_2$, is a good estimate of the
interaction between the configurations $U_1$ and $U_2$. 
However, even if $U_1$ and $U_2$ have a common symmetry, $U$
typically has this symmetry slightly broken. This is
connected with the principal difficulty of the product ansatz, that
the field $U$ changes if the product is taken in the opposite order.
The product ansatz is not useful for approximating true, minimal
energy, Skyrmion solutions.   

The rational map ansatz is a good method to approximate Skyrmion solutions.
Here the idea is to
separate the angular from the radial dependence of the Skyrme field,
and to use a complex (Riemann sphere) coordinate 
$z = \tan \frac{\theta}{2} \, e^{i\phi}$ in place of the
usual spherical polar coordinates $\theta$ and $\phi$.

The Skyrme field is constructed from a rational function of $z$
\be
R(z) = \frac{p(z)}{q(z)} \,,
\ee
where $p$ and $q$ are polynomials in $z$ with no common root, 
together with a radial profile 
function $f(r)$ satisfying $f(0) = \pi$ and $f(\infty) = 0$. One should
think of $R$ as a smooth map from a 2-sphere in space (at a given
radius) to a 2-sphere in the target $SU(2)$ (at a given distance from
the identity). By standard stereographic projection, an image point
$R$ can be expressed as a Cartesian unit vector
\be
\hat{\bf n}_R = \frac{1}{1 + |R|^2}(R + \bar{R}, i(R -
\bar{R}), 1 - |R|^2) \,.
\ee
The ansatz for the Skyrme field is then
\be
U(r,z) = \exp(if(r)\hat{\bf n}_{R(z)}\cdot\pauli) \,,
\label{ratansatz}
\ee
or equivalently
\be
U(r,z) = \cos f(r){ 1} + i \sin f(r) \hat{\bf n}_{R(z)}\cdot\pauli \,.
\ee
This should be thought of as a suspension of the map $R:S^2 \to S^2$
to produce a map $U:\bR^3 \to S^3$, the suspension points being the
origin and infinity in $\bR^3$, which are mapped to $U = -1$ and
$U = 1$, respectively.

The topological degree of the rational map $R:S^2 \to S^2$ is the
higher of the algebraic degrees of $p$ and $q$, and it is not
difficult to show that the baryon number, $B$, of the Skyrme field is equal
to this.

An important feature of the rational map ansatz is that when one
substitutes it into the Skyrme energy function (\ref{skyenergy}), 
the angular and radial parts decouple. To minimize the energy (for 
given $B$), it is sufficient to first minimize a certain angular 
integral $\I$ that only depends on 
the coefficients occurring in the rational map, and then to solve an
ordinary differential equation for $f(r)$. The coefficients of the
ODE depend on the rational map, but only through its degree $B$, and the
value of $\I$. Both steps can be carried out numerically,
efficiently and accurately. Optimal rational maps,
and the associated profile functions, have been found for many values
of $B$ \cite{BS3}. The optimal map often has a high degree of symmetry, and
the Skyrme field $U$ inherits this symmetry exactly. Considerable
effort has been devoted to understanding the classes of rational maps
with cyclic, dihedral, and platonic symmetries.

The optimized fields within the rational map ansatz can either be
regarded as good approximations to exact solutions of the Skyrme
equations (which almost always have the same symmetry), or they can be
used as a starting point for a numerical relaxation to exact Skyrmion
solutions. For the polyhedral shell Skyrmions, at zero pion mass, the 
rational map ansatz is already a very good approximation.

For the new solutions we discuss in this paper, with an
$\alpha$-particle cluster
structure, the original rational map ansatz is only occasionally
useful. More helpful is the double rational map generalisation.
This uses two rational maps $R^{\rm in}(z)$ and 
$R^{\rm out}(z)$, with a profile function $f(r)$ satisfying  
$f(0) = 2\pi$ and $f(\infty) = 0$. It is assumed that $f$ decreases
monotonically as $r$ increases, passing through $\pi$ at a radius $r_0$.
The ansatz for the Skyrme field is again (\ref{ratansatz}), with the
understanding that for $r < r_0$, $R(z) = R^{\rm in}(z)$, and for 
$r > r_0$, $R(z) = R^{\rm out}(z)$. Notice that now $U = 1$ both at the
origin and at spatial infinity, and $U = -1$ on the sphere $r = r_0$.
It can be shown that the total baryon number is the sum of the degrees
of the maps $R^{\rm in}$ and $R^{\rm out}$. The ansatz is optimized by
adjusting the coefficients of both maps, allowing variations of $r_0$,
and solving for $f(r)$. All this is quite hard, but easier if 
$R^{\rm in}$ and $R^{\rm out}$ share a substantial symmetry. 

In \cite{MP}, it was hoped that by using this double rational map
ansatz (or its generalisation),
and then relaxing the field, new Skyrmions would be found. However,
the analysis was restricted to zero pion mass, and baryon numbers no
higher than 14. The results were disappointing, since only solutions
of relatively high energy were found. Our present work shows
that the ansatz
is more useful in providing initial conditions
when the pion mass is positive, and when one considers 
higher baryon numbers. 

\section*{Acknowledgements}
Many thanks to Steffen Krusch for useful discussions.
This work was supported by the PPARC special programme
grant ``Classical Lattice Field Theory''. 
The parallel computations were performed on COSMOS at the National
Cosmology Supercomputing Centre in Cambridge.

\end{document}